\documentclass[amssymb,useAMS,prd,aps,amsmath,superscriptaddress,nofootinbib]{revtex4}
\usepackage{color}
\usepackage{pslatex}
\usepackage{psfrag}
\usepackage{graphicx}
\usepackage{array}
\begin{document}

\title{Radio observations of the Galactic Centre and the Coma cluster as a probe of light dark matter self-annihilations and decay}

\author{C\'eline B\oe hm}
\affiliation{LAPTH, CNRS, 9 chemin de Bellevue - BP 110,
  74941 Annecy-Le-Vieux, France.}

\author{Joseph Silk}
\affiliation{Astrophysics department, DWB building, Keble Road, OX1 3RH Oxford}

\author{Torsten En$\beta$lin}
\affiliation{Max-Planck-Insitut f{\"u}r Astrophysik
Karl-Schwarzschild-Str. 1
85741 Garching
Germany}

\date{today}

\begin{abstract}
We update our earlier  calculations of gamma ray and radio observational constraints on annihilations of dark matter particles lighter than 10 GeV.
We predict the synchrotron spectrum as well as the morphology of the radio emission associated with light decaying and annihilating dark matter candidates in both the Coma cluster and the  Galactic Centre. Our new results basically confirm our previous findings: synchrotron emission in the very inner part of the Milky Way  constrains or even excludes dark matter candidates if the magnetic field is larger than 50 $\mu$ G. In fact, our  results suggest that light annihilating candidates must have a S-wave suppressed pair annihilation cross section into electrons (or the branching ratio into electron positron  must be small). If dark matter is decaying, it must have a life time that is larger than $t = 3. \ 10^{25} \rm{s}$. Therefore, radio emission should always be considered when one proposes a ``light'' dark matter candidate.  
\end{abstract}
\maketitle

\section{Introduction}

Recently, many papers have entertained the idea of light dark matter particles in the 1-10 GeV range \cite{Kuflik:2010ah,Fitzpatrick:2010em,Andreas:2008xy,Asano:2009kj,Hisano:2009xv,Cheung:2009wb,Hooper:2008au,Bottino:2009km,Feldman:2010ke,Bae:2010hr,Mambrini:2010dq,Cline:2010kv}. The increasing interest in  such candidates stems mostly from recent claims by three dark matter direct detection experiments (DAMA/LIBRA \cite{Bernabei:2008yi,Bernabei:2010mq}, CDMS \cite{Ahmed:2009zw} and CoGeNT \cite{Aalseth:2010vx})  of a possible dark matter signal in their detectors. Although the dark matter mass initially favoured by CDMS was in the 40-60 GeV range, it was found that CDMS data are also compatible with $\sim$ 10 GeV candidates (the range  of masses required to account for  CoGeNT events), thus raising the exciting possibility that  CDMS, CoGeNT and DAMA/LIBRA have found evidence for relatively light dark matter particles.

In reality, the situation is more complex than it seems. Both XENON 10 and XENON 100 experiments place  severe constraints on low mass dark matter particles (even though there is extensive  discussion of large uncertainties in  the scintillation efficiency adopted by XENON). Also, the cross sections favoured by DAMA/LIBRA seem larger than those required to explain CoGeNT. 
To make the situation even more confusing, CRESST has announced, as preliminary results, about 30 (not yet identified) events in the oxygen band \cite{idm} while EDELWEISS identified three potential candidate events in the dark matter region (with an energy resolution greater than 20 keV, i.e. larger than that of CoGeNT) \cite{ioa},  all of which remain to be analysed carefully.

At present, whether or not  these experiments have detected a dark matter signal for the first time remains a highly speculative question. But, from the theoretical point of view, the confirmation of a dark matter signal would unveil a mass range that remains relatively unexplored. Of course, the idea of light dark matter particles is not new. 
It  was investigated in Ref.~\cite{2001PhRvD..63b3001M,1990PhRvD..41.3080F,bfs,bens}. 
The main conclusion from these three studies is that viable dark matter particles lighter than the GeV scale should have S-wave suppressed pair annihilation cross sections into electron-positron pairs, based on CMB, BBN and gamma ray observations respectively. However, the astrophysical implications of the specific range between 1 and 10 GeV was also studied in Ref.\cite{bens}. Similar conclusions were obtained as for the lower mass range, i.e. particles in the  1-10 GeV range should also have  S-wave suppressed pair annihilations in order to not overproduce the \textit{radio emission} in the Galactic Centre. It was shown, in addition, that such particles would produce  a visible signal in the Coma cluster  of galaxies if the dark matter mass was about 10 GeV. Therefore, the conclusion that had emerged at that time was that radio emission should always be estimated when one considers the types of candidates that have been proposed to explain the data from CoGeNT and CDMS,  since the synchrotron spectrum can constrain their interactions.

Particles with masses within  the mass range 1-10 GeV have since  been studied from the particle physics perspective, including a supersymmetric context (see Ref.\cite{Bottino:2002ry,Hooper:2002nq,Belanger:2003wb}). However, the radio constraints were disregarded, even though the requirement of a S-wave suppressed pair annihilation cross section has strong implications. This condition could  indeed imply e.g. the introduction of a light $Z'$ (see Ref\cite{bens,bf}) or even offer a link between light particles and neutrino masses \cite{Boehm:2008zz}. The $Z'$ couplings to baryonic matter would have to be small \cite{Boehm:2004uq} but this would be compatible with direct detection limits on spin-independent cross sections. 

In absence of a new light gauge boson (and disregarding the radio constraints that were derived in Ref.~\cite{bens}), the question of the relic density of light thermal dark matter particles is a rather difficult issue. One possibility is to assume that there are significant annihilations into neutrinos  \cite{Boehm:2008zz}. Another solution is to assume e.g. a freeze-in scenario \cite{Hall:2009bx} or some dark matter/anti dark matter number density asymmetry. One could also postulate that the candidates under study represent only a small fraction of the dark matter relic density (i.e. that there are several species of dark matter) \cite{Boehm:2003ha}. Or one could simply abandon the thermal relics paradigm. Alternatively, one could assume that annihilations of light dark matter particles proceed
through a Higgs exchange (as in Ref.~\cite{Bottino:2002ry,Hooper:2002nq,Belanger:2003wb}). However, the recent measurement of the $B_s \rightarrow \mu \mu$ process now severely constrains such a scenario \cite{Feldman:2010ke}.

Building a  light thermal annihilating dark matter model from the particle physics point of view is therefore not an easy task. But, here, we demonstrate that this may be even more difficult since synchrotron emission provides an additional constraint that cannot be disregarded. 

To illustrate our argument, we will consider both decaying and annihilating dark matter candidates. When we will include the electron propagation, we will focus on two values of the dark matter mass in particular: namely 5 and 10 GeV. To be as model independent as possible, we disregard the production of $\mu$,$\tau$ and light quarks by the dark matter. However one can easily repeat these calculations by adding a function $N(E)$ that takes into account the multiplicity of electrons for each value of the energy if dark matter produces other particles than electrons.

In Sec.\ref{wodiffusion}, we extend the calculations that we published in Ref.~\cite{bens} and give an analytical expression for the flux that is  expected for both decaying and annihilating candidates  in the Galactic Centre. This analytical expression is obtained by neglecting spatial diffusion and must be used with caution. However, it does enable us to obtain a rough estimate of whether a candidate is likely to be excluded or not. In Sec.\ref{withdiffusion},  we use a semi-analytical model to include spatial diffusion. We predict the synchrotron spectrum  in the inner part of the galactic centre associated with  dark matter annihilations or decays into electrons and also show maps of the morphology of the emission for a magnetic field of 100 $\mu$G. We present our  conclusions in the final section.

\section{Synchrotron predictions in absence of electron spatial diffusion \label{wodiffusion}}

Our goal in this section is to provide a quick estimate of the synchrotron emission due to the presence of dark matter particles in the galaxy and to estimate the synchrotron flux in the Coma cluster and the galactic centre due to dark matter. With this aim in mind, we neglect spatial diffusion of the electrons. The emissivity at a frequency $\nu$ (corresponding to electrons with an energy $E(\nu) = \sqrt{\frac{\nu}{ 16 \ \rm{MHz}}  \ \frac{1}{B_{\mu \rm{G}}}}$ GeV) and at a position $r$  reads:
$$\epsilon_{\nu}(r) = n(E,r)   \times \frac{E}{\nu} \times P(E) $$ 
where $P(E)$ is the power radiated by the synchrotron emission: 
$$P(E) = \frac{1}{6 \pi} \ \sigma_T \ c \ \beta^2 \ \gamma^2 \ B^2,$$ and  $n(E,r)$ is the number of electrons with an energy E at the position r.
We can then define the observable surface brightness, as the integral over the line-of-sight of the emissivity:
$$I_{\nu}(r,\psi)  = \frac{1}{(4 \pi)} \ \int ds(r,\psi) \ \epsilon_{\nu}(r)$$
and the flux $F_{\nu}$ within an annulus $\Omega_{\psi}$ as:
$$ F_{\nu}(r) = \int d\Omega_{\psi} \ I_{\nu}(r,\psi).$$
In the case of dark matter annihilation or decay, $ n(E,r)$  is given by:
$$n(E,r) =   \ \frac{1}{b(E)} \frac{N_e}{\eta_n} \ Q_n \left(\frac{\rho}{m_{dm}}\right)^n \ \theta\left(E- \frac{n \ m_{dm}}{2}\right),$$
where $b(E)$ represent the energy losses and $N_e$ is the number of electron and positron emitted during the dark matter decay or annihilation (that is $N_e=2$ in the present case). 
Hence, the radio flux that is expected from dark matter annihilation or decay is  given by the following expression:

\begin{eqnarray}
F_{\nu}(r) &\sim& 1.1 \ 10^8 \ Jy  \times \frac{N_e}{\eta_n}  \times \left(\frac{Q_n}{3.10^{-26} \rm{units}}\right) \times \left(\frac{m_{\rm{dm}}}{\rm{GeV}}\right)^{-n} \times {\cal{F}}_n \times \left(\frac{\nu}{GHz}\right)^{-1/2}  \nonumber \\
&&\hspace{1cm} \times \left(\frac{B}{\mu G}\right)^{-1/2}  \times \left(\frac{\rho_0}{1. \rm{GeV/cm^3}}\right)^n \times \left(\frac{P(E)}{b(E)}\right)_{\vert_{E(\nu)}}.
\label{eq1}
\end{eqnarray}

Our convention is the following: $n=1$ denotes decaying dark matter particles, $n=2$ annihilating dark matter particles and the function 
$${\cal{F}}_n =  \ \left[\frac{1}{4 \pi} \int d \Omega_{\psi} \int \ \  \frac{d s(r,\psi)}{\rm{kpc}} \ \ g(r)^n \right]$$ represents the integral over the line of sight of the normalized dark matter halo profile ($g(r) = \rho(r)/\rho_0$) in an annulus defined by $d \Omega_{\psi}$. 
Note that Eq.\ref{eq1} is only valid in the case of  the monochromatic approximation, where one frequency corresponds to one electron energy, and vice versa.

In this expression, $\eta_n = S \times \left(\frac{\rho_{\chi}}{\rho_{dm}}\right)$ where $\rho_{\chi}$ denotes the energy density of the dark matter candidate and $\rho_{dm}$ denotes the dark matter energy density in the halo. $S$ is a symmetry factor. When the dark matter is annihilating and made of self-conjugate particles (that is, if it is a Majorana or a real scalar for example), $\rho_{\chi} = \rho_{dm}$ but $S=1/2$. Hence $\eta_{n=2}=2$ in the case of Majorana particles. In the case of  annihilating non self-conjugate particles (e.g. for a Dirac particle or complex scalar), $\rho_{\chi} = \rho_{dm}/2$ and $S=1$. Hence $\eta_{n=2}=4$. On the other hand, for decaying particle, $\eta_{n=1}=1$. 

The term $Q_n$ denotes either the decay rate (in $\rm{s^{-1}}$) or the annihilation cross section (in $\rm{cm^3/s}$), depending on whether dark matter is decaying  or annihilating respectively. The integration over the line of sight and solid angle can be done analytically \cite{Lavalle:2009fu} if one considers the inner part of the galaxy where $r<r_s$ and  assumes, in addition, that the  profile within $r<r_s$ is well approximated  by $g(r)= (r/r_s)^{-\gamma}$. In this case, one can rewrite ${\cal{F}}_n$ in the following form:
\begin{eqnarray}
{\cal{F}}_n &=& \frac{r_s^3}{2 \ d^2 } 
\times   \int^{\alpha_{\psi}}_0 d \alpha \ \ \alpha^{2 -n \, \gamma}  \ \int^{\sqrt{1-\alpha^2}/\alpha}_{\alpha} \ dv \frac{1}{(v^2+1)^{n \gamma/2}} \nonumber \\
&=& \frac{r_s^3}{2 \ d^2 }  \times G_n 
\end{eqnarray}
where $ \alpha_{\psi} =  \ d \psi/r_s$ ($d$ the distance to the object from the observer). When $\gamma=1$ (that is, when $\rho_{dm} \propto r^{-1}$), we then obtain that $G_n$ is equal to:  

\begin{eqnarray}
G_{n=1} &\simeq&  \alpha_{\psi}^2 \ \ln\left(\frac{2}{\alpha_{\psi}}\right) \nonumber \\
G_{n=2} &\simeq& \frac{\pi}{2} \ \alpha_{\psi} \nonumber \\
\nonumber
\end{eqnarray}

for decaying and annihilating DM respectively, leading to:

\begin{eqnarray}
{\cal{F}}_{n=1} &\simeq&  \frac{r_s}{2} \times \left(\frac{\pi}{180}\right)^2 \times \psi^2 \times \ln\left(\frac{360 \ r_s}{ d \ \pi \ \psi}\right) \label{F1} \\
{\cal{F}}_{n=2} &\simeq& \frac{r_s^2}{d} \times \frac{  \pi^2}{(4\times 180)} \times \psi \label{F2} \\
\nonumber
\end{eqnarray}
with $\psi$ expressed in degrees. The same expressions (eqs.~\ref{F1} and \ref{F2}) can be used for Coma, provided that $\psi \lesssim 0.1$ deg.  For larger values of $\psi$, the condition $d \alpha < r_s$ is no longer valid and one has to integrate numerically the profile (in which case, one does not need to assume $r<r_s$ anymore).

\subsection{Cluster of galaxies} 

Let us now consider a dark matter candidate with a mass $m_{dm}$. Electrons are injected with an energy $E_{inj}= \frac{n}{2} \times m_{dm}$, which corresponds to $E_{inj}= m_{dm}/2$ for decaying dark matter and $E_{inj} = m_{\rm{dm}}$
for annihilating dark matter. The electrons thus produced lose their energy very quickly down to an energy $E$ which corresponds to a specific value of the  propagation length. In this model, one therefore expects synchrotron emission at all frequencies smaller than the frequency $\nu_{max}$, corresponding to:

$$\nu_{max} = 16 \rm{MHz} \times \left(\frac{n}{2}\right)^2 \times \left(\frac{m_{dm}}{ \rm{GeV}}\right)^2 \times  \left(\frac{B}{\mu G}\right).$$

Radio observations of the Coma cluster range from  30.9 MHz to 4.8 Ghz frequencies. In particular, measurements have been done at 30.9, 43, 73.8, 151, 326, 408, 430, 608.5, 1380, 1400, 2700, 4850 Mhz. Assuming a magnetic field value of 1 $\mu G$,  candidates with a mass of $1.39$ and $17.36$ GeV generate radio emission at 30.9 MHz and 4.8 GHz respectively. Hence, using radio observations of the Coma cluster (A1656) could actually be relevant in light of DAMA/LIBRA, CDMS, CoGeNT and possibly CRESST findings. 

In what follows, we consider an annulus of $1$ degree (corresponding to a solid angle of $\sim 10^{-3}$ sr). We perform the integration numerically. Also, we take $d = 100$ Mpc, $r_s = 400$ kpc, $\rho_0  = 4.4 \ 10^{-2} \ \rm{GeV/cm^3}$ and assume that the number density of the gas  is about $n_b^{th} = 3 \ 10^{-3} \rm{cm^{-3}}$ \cite{bens}. Since Coulomb and bremsstrahlung losses (which are particularly important at low energy) are proportional to $n_b^{th}$, the synchrotron emission at low frequency is sensitive to $n_b^{th}$.

\begin{figure}[h]
	\centering	
	\includegraphics[width=8cm]{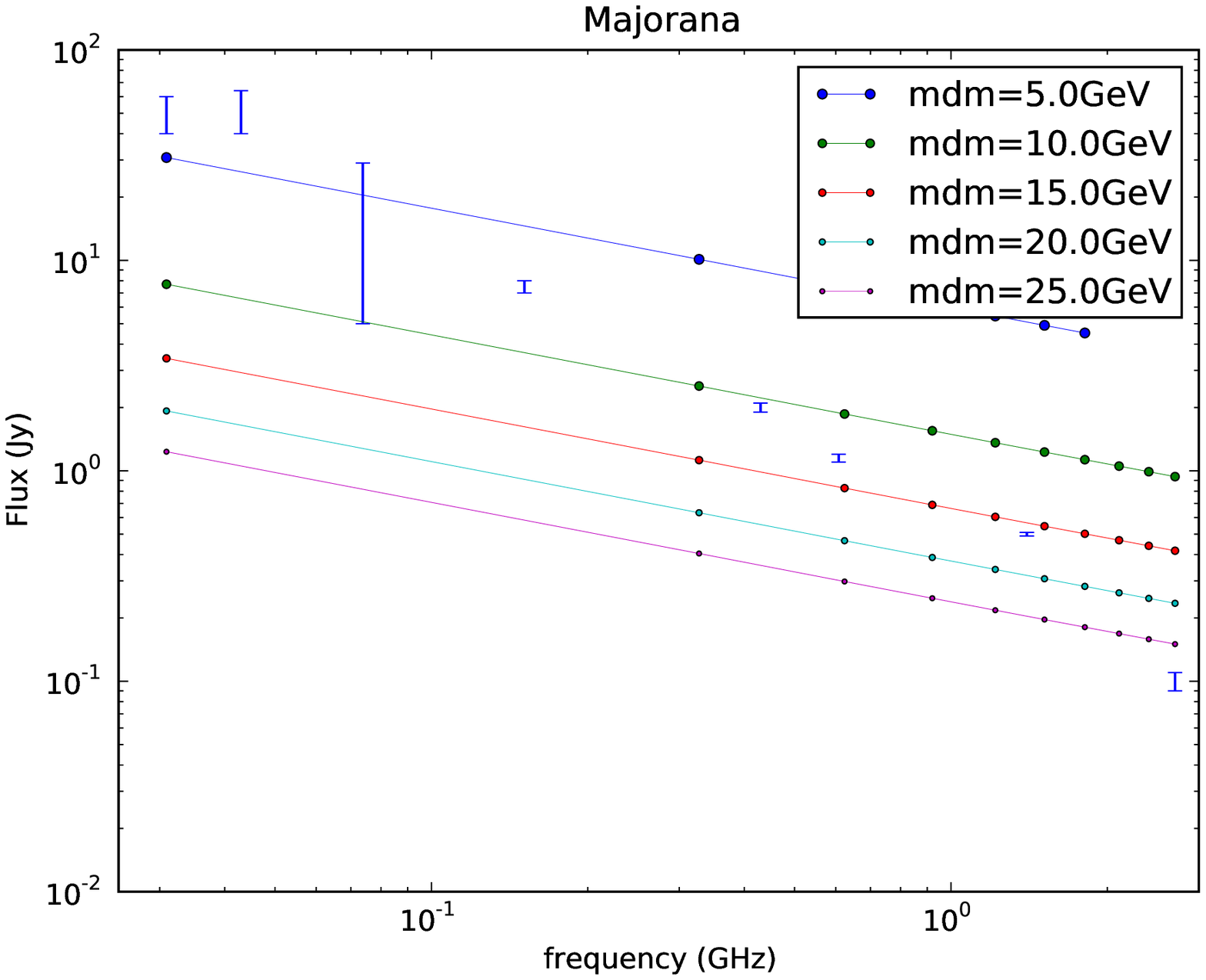} 
		\includegraphics[width=8cm]{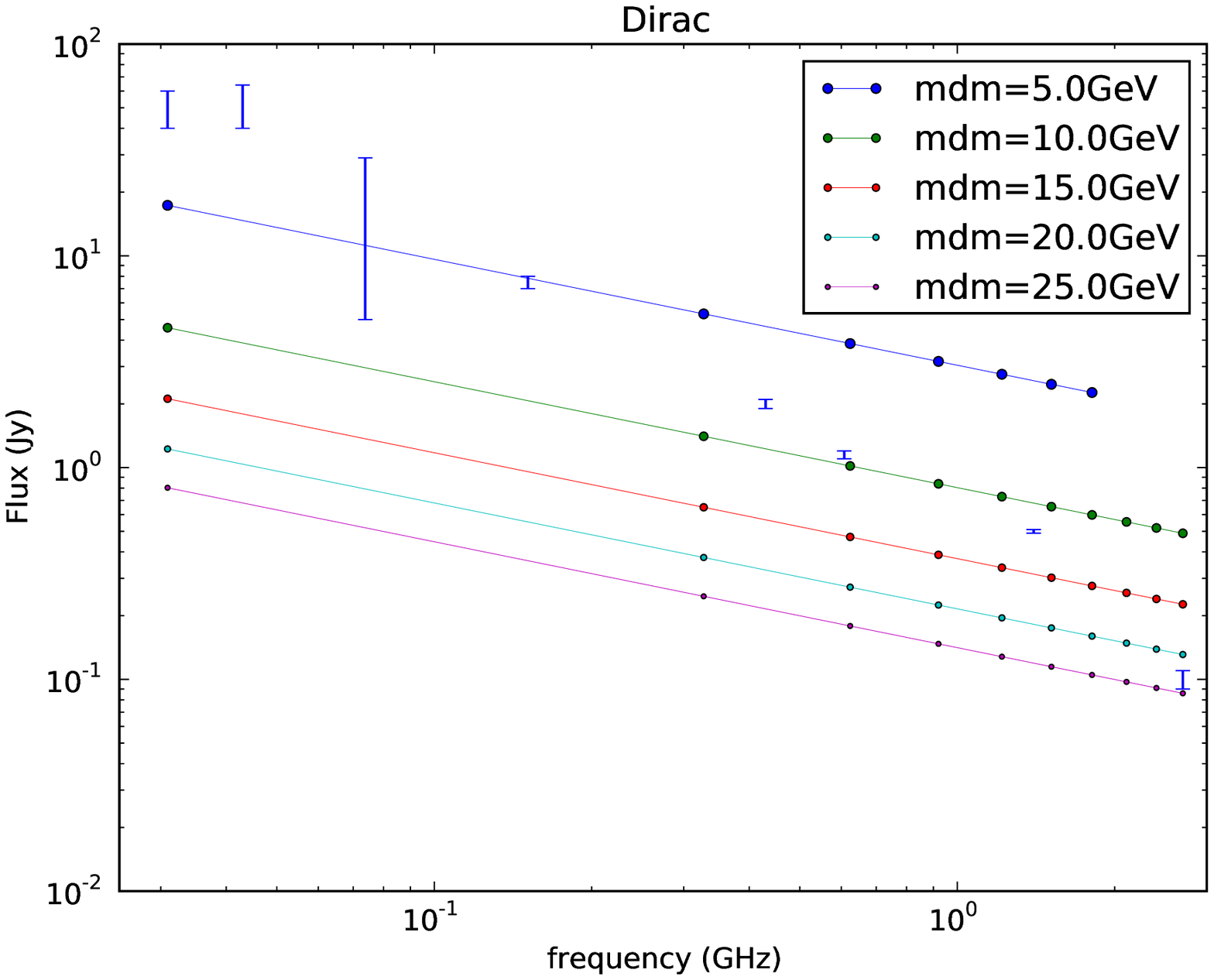}
	\caption{Flux expected from the annihilation of Majorana and Dirac dark matter particles into electron-positron in the Coma cluster with masses ranging from 1 to 25 GeV. Predictions for Majorana particles are displayed in the left panel while those for Dirac particles are shown in the right panel. We have taken a magnetic field of $B=4.7 \mu G$, an annihilation cross section of $\sigma v = 3 \ 10^{-26} \ \rm{cm^3 s^{-1}}$ and a NFW profile ($\gamma=1$). The measurements are represented as error bars on this figure. }
	\label{fig:eps_a_nonboosted_coma}
\end{figure}

To make the comparison easier, we have chosen the same value of the magnetic field as in Ref.\cite{Colafrancesco:2010kx}, that is $B=4.7 \mu$G. However, we took the canonical value of the dark matter pair annihilation cross section (that is $\sigma v = 3 \ 10^{-26} \ \rm{cm^3/s}$) while Ref.\cite{Colafrancesco:2010kx} has considered larger values of the cross section.  One can nevertheless easily rescale our results accordingly to take into account a boost factor (whether it originates from particle physics or astrophysics).  Our results are displayed in Fig.~ \ref{fig:eps_a_nonboosted_coma}.

The fact that we have considered monochromatic electrons at injection (instead of a spectrum given by $b$ or $\tau$ cascades as in Ref.\cite{Colafrancesco:2010kx}) obviously leads to different predictions with respect to  Ref.\cite{Colafrancesco:2010kx}. However, the orders of magnitude (after rescaling) are consistent with the results displayed in Ref.\cite{Colafrancesco:2010kx}) at small frequencies. As one can see, with our value of the cross section and normalisation of the dark matter halo profile, one can easily rule out annihilating particles with a mass comprise in 1-10 GeV, using radio observations at $\nu <1$ GHz and up to 20-25 GeV using $\nu <3$ GHz. However, these results may be weakened if electrons mostly comes from the decay of $\mu,\tau$ or light quarks and they may change with the value of the magnetic field. Hence, one can simply conclude from this exercice that radio emission in the Coma cluster should always be estimated properly for light annihilating candidates if their S-wave pair annihilation cross section is not overly suppressed.

\begin{figure}[h]
	\centering		\includegraphics[width=8cm]{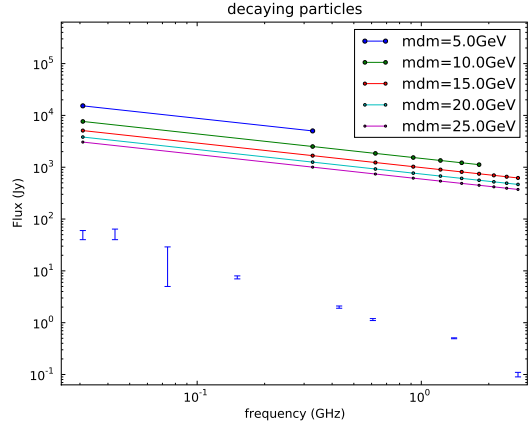}
	\caption{Synchrotron emission from the decay of dark matter particles into a pair electron-positron in the Coma cluster. We have chosen $B=4.7 \mu G$, a (two-body) decay rate into electron-positron of $\Gamma = 3 \ 10^{-26} \ \rm{s^{-1}}$ and a NFW profile ($\gamma=1$).}
	\label{fig:eps_d_nonboosted_coma}
\end{figure}

For decaying dark matter, our results are displayed in Fig.~\ref{fig:eps_d_nonboosted_coma}. The constraints are much more stringent than in the annihilating case because of the integration over the line of sight of the dark matter halo profile. Light (1-25 GeV) dark matter particles must have a (two-body) decay rate into electrons and positrons that is smaller than $10^{-27} \ \rm{s}$   to not overproduce the radio emission in the Coma cluster. Assuming that there is no other decay channel, this translates into a lifetime that is actually larger than the lifetimes generally constrained by gamma ray observations. Hence, synchrotron estimate is actually a very powerful tool for constraining light decaying dark matter particles.

\subsection{Milky Way} 

We can now apply our expression of the flux to light dark matter particles that would annihilate or decay (or both) in the Milky Way. In particular, we are interested in the very inner part, corresponding to very small value of the solid angle ($10^{-5} sr$ centred on the galactic centre). 

The dependence of the flux upon the magnetic field arises only from the energy losses and the energy. However, when the magnetic field becomes too large, the synchroton dominates over all the other losses and the factor $P(E)/b(E)$ tends to unity. For smaller values of the magnetic field (such as 1 $\mu$ G), the factor $P(E)/b(E)$ is much smaller than unity (still assuming $n_b^{th}=1 \ \rm{cm^{-3}}$). Hence, depending on the value of the magnetic field, the radio flux is also suppressed by the ratio of the synchrotron power radiated to the energy losses.

\begin{figure}[h]
	\centering	
	\includegraphics[width=8cm]{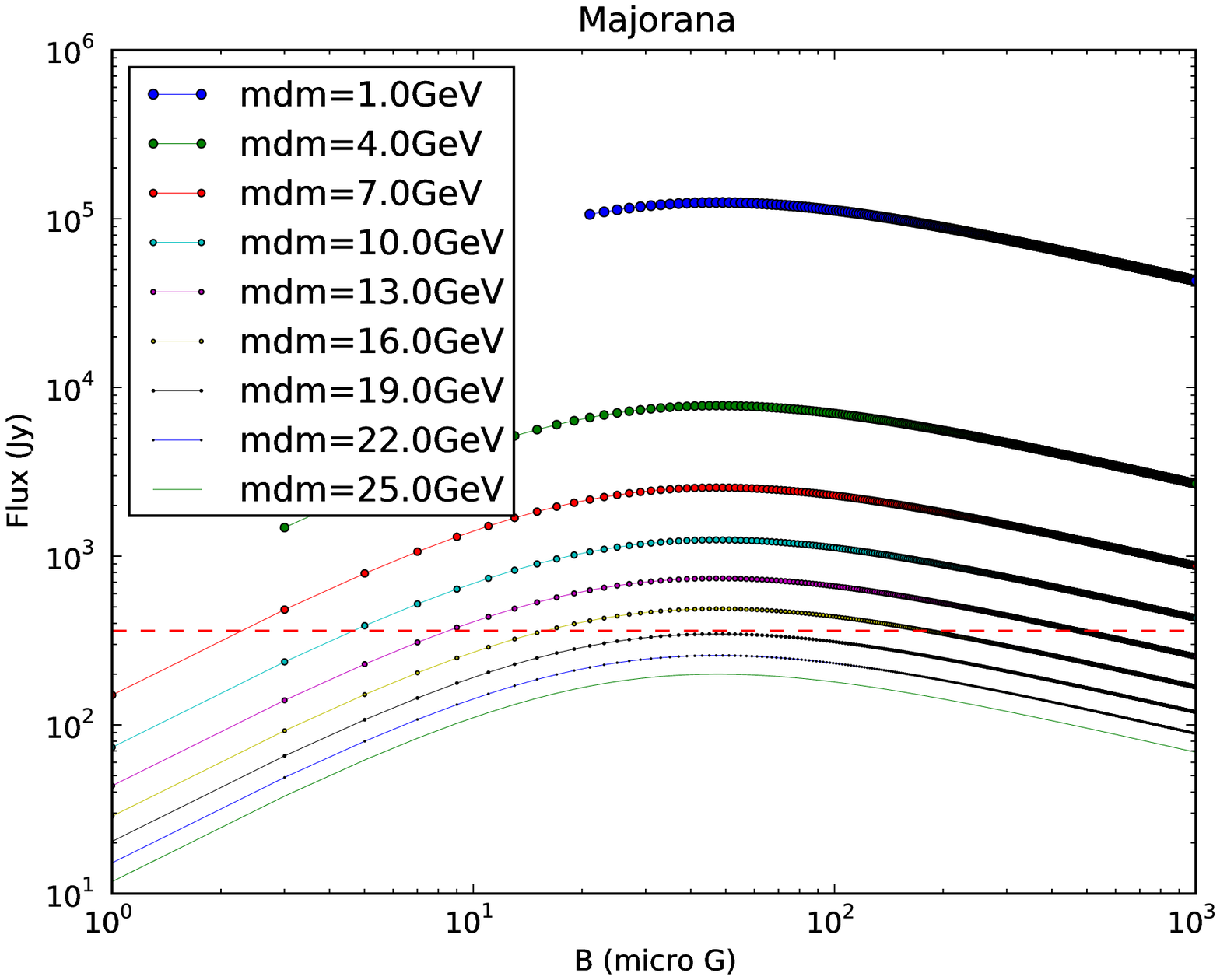}
		\includegraphics[width=8cm]{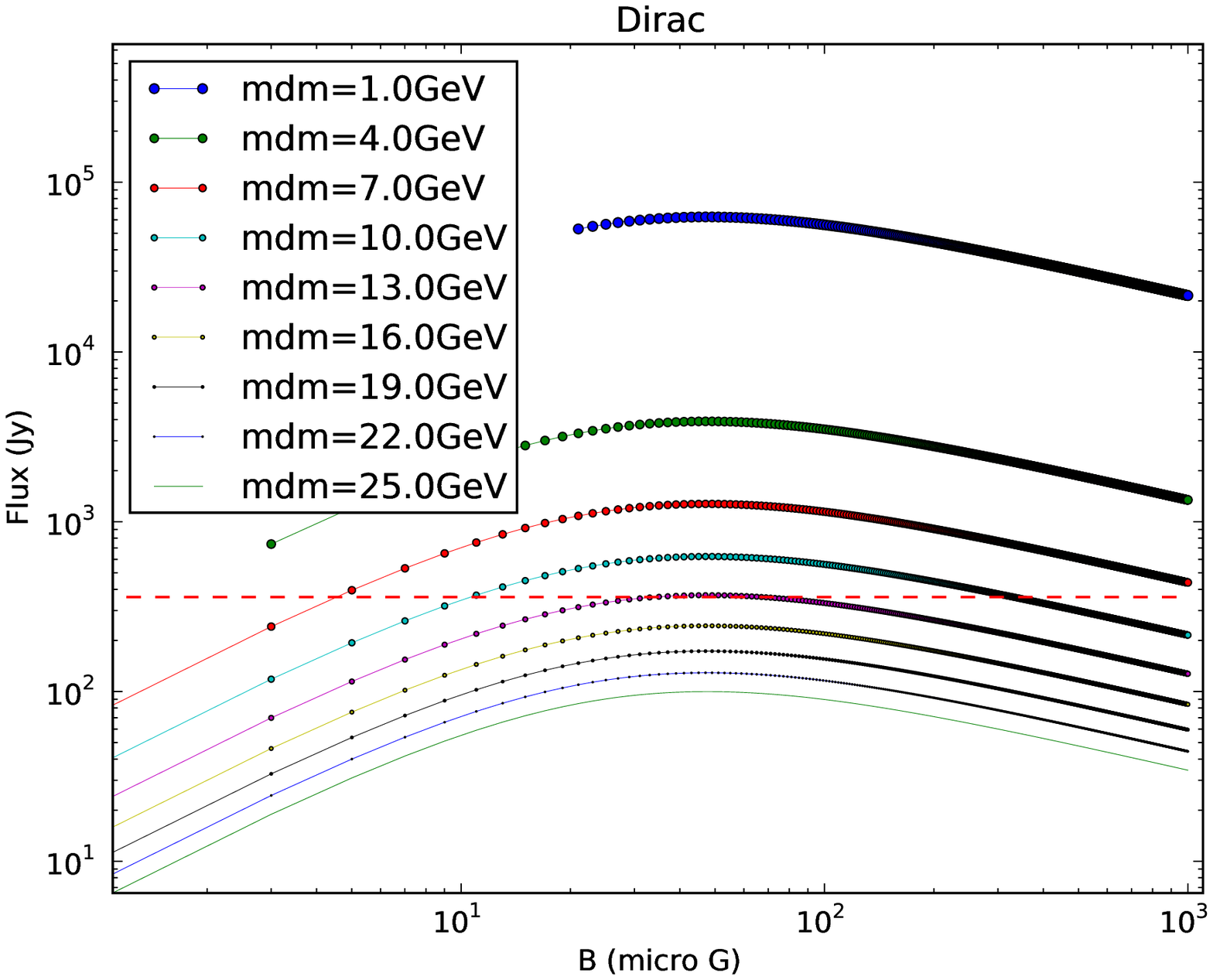}
	\caption{Flux expected in the Milky Way  at 330 MHz for thermal Majorana and Dirac annihilating dark matter candidates with masses equal to (from top to bottom) 1, 4, 7, 10, 13, 16, 19, 22, 25 GeV and a solid angle  of $\sim 10^{-5}$ sr. We have taken an annihilation cross section of $\sigma v = 3 \ 10^{-26} \ \rm{cm^3 s^{-1}}$ and a NFW profile ($\gamma=1$).}
\label{fig:epsnua}
\end{figure}

In Fig.\ref{fig:epsnua}, we show the radio flux that is expected in the Milky Way from dark matter annihilations as a function of the magnetic field for different values of the dark matter mass.
We have assumed a S-wave annihilation cross section. Besides, we use for comparison the emission of SgrA at 330 MHz, which is about 360 Jy (see dotted line on Fig.~\ref{fig:epsnua}). As one can see in this figure, thermal annihilating dark matter particles (with a velocity-independent annihilation cross section) always overproduce the radio emission in the galactic centre if their mass lies in between 1 and $\sim$ 7-10 GeV, whatever the value of the magnetic field. Yet, the contribution for both self-conjugate and non-conjugate dark matter candidate with a mass greater than 10 GeV and smaller than 25 GeV can remain important whatever the value of the magnetic field and sometimes, in fact, large enough to indicate that the candidate is likely to be excluded. Indeed, all the  candidates with $m_{dm} \in [7-10,25]$ GeV would contribute significantly (if not totally) to the inner radio emission of the galaxy if $B \in [20,100] \mu G$ in the Majorana case. The same remark is true for $m_{dm} \in [1,20]$ GeVin the case of Dirac particles.

Note that radio emission for 1 to 4 GeV candidates require that the magnetic field is larger than 20 and 3$\mu G$ respectively. These values are supported by the lower limit of 50 $\mu$G which was found in the very inner part of the galactic centre (see  Ref.\cite{Crocker:2010xc}). However, this limit should be used with caution since the radio contribution from dark matter decay and/or annihilations may actually affect the value of the magnetic field that is  inferred from observations. Yet, this should only be relevant when $m_{dm}$ is large enough (that is, when the radio contribution from dark matter does not exceed the measured flux in the inner galaxy by several orders of magnitude). Since synchroton emission for 1-10 GeV particles is very large, light dark matter candidates with a S-wave annihilation cross section equal to the canonical value (i.e. $3 \ 10^{-26} \ \rm{cm^3/s}$) are likely to be excluded by radio observations. We came to the same conclusion by considering a solid angle of $\sim 10^{-3}$ sr (centred on the galactic centre) since the observed flux in this region of the sky does not exceed $10^4$Jy.

Therefore, as we had already found in \cite{bens},  if  dark matter is made of a light thermal annihilating candidate (i.e. with a pair annihilation cross section of $3. \ 10^{-26} \rm{cm^3/s}$ in the primordial Universe to explain the observed dark matter abundance), it must either be heavier than 7-10 GeV  (remembering that the range up to 20 GeV is likely to be constrained too) or rely on some dark matter velocity-dependent process to be compatible with radio observation of the galactic centre.

This severely constrains the models which have been proposed to explain the results from DAMA/LIBRA, CoGeNT and CDMS. For example, we found that for $m_{dm}=$ 8 GeV (which corresponds to the type of  mass which was suggested to explain the results from CoGeNT) and a magnetic field in the range 20-100 $\mu$G, the S-wave annihilation cross section into electrons must be suppressed by a factor 3 to 10 (for self-conjugate dark matter) and a factor 3 (for non self-conjugate dark matter) with respect to the canonical value $3. \ 10^{-26} \ \rm{cm^3/s}$ in order for the model to be acceptable. 

\begin{figure}
	\centering	
	\includegraphics[width=8cm]{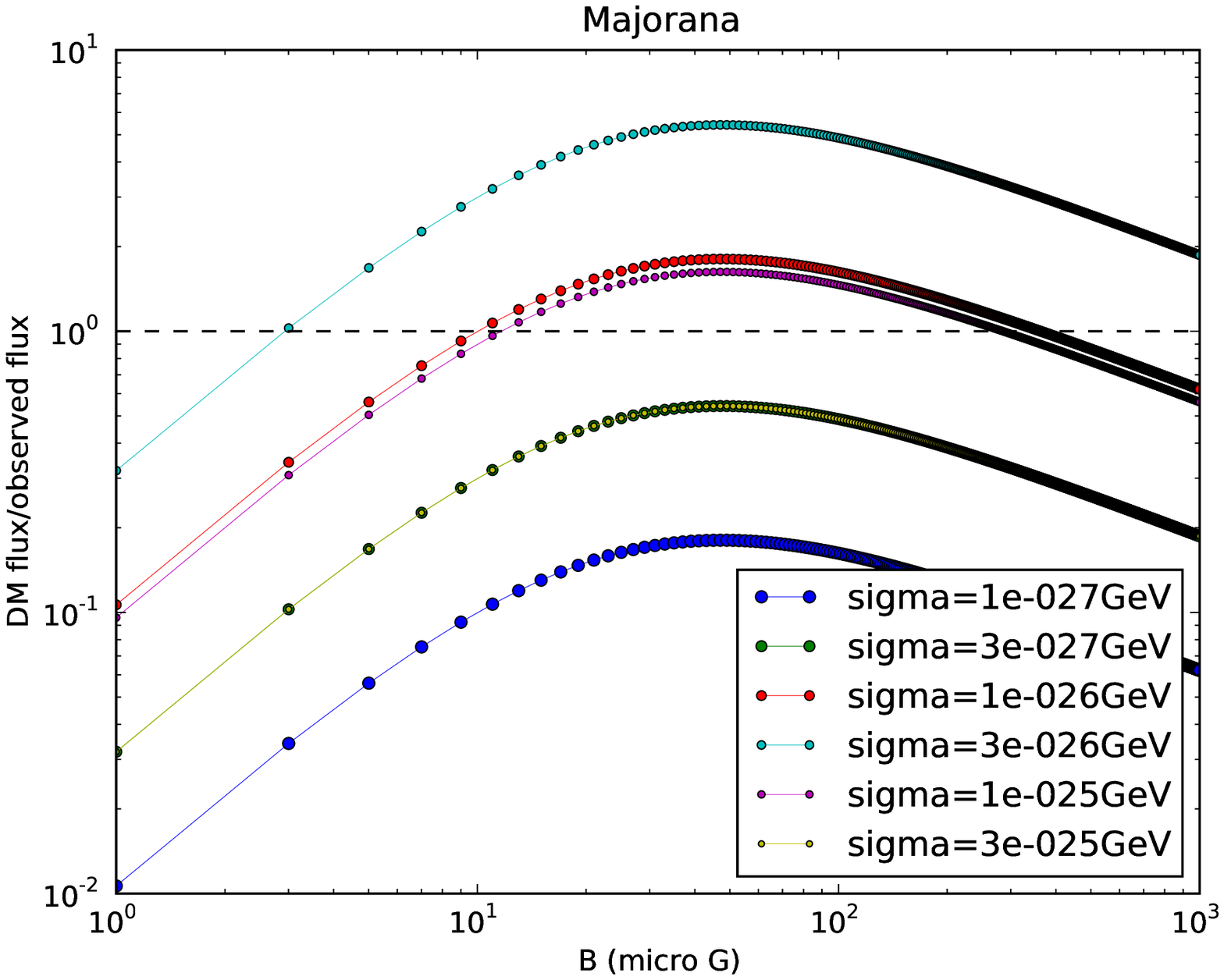}
	\includegraphics[width=8cm]{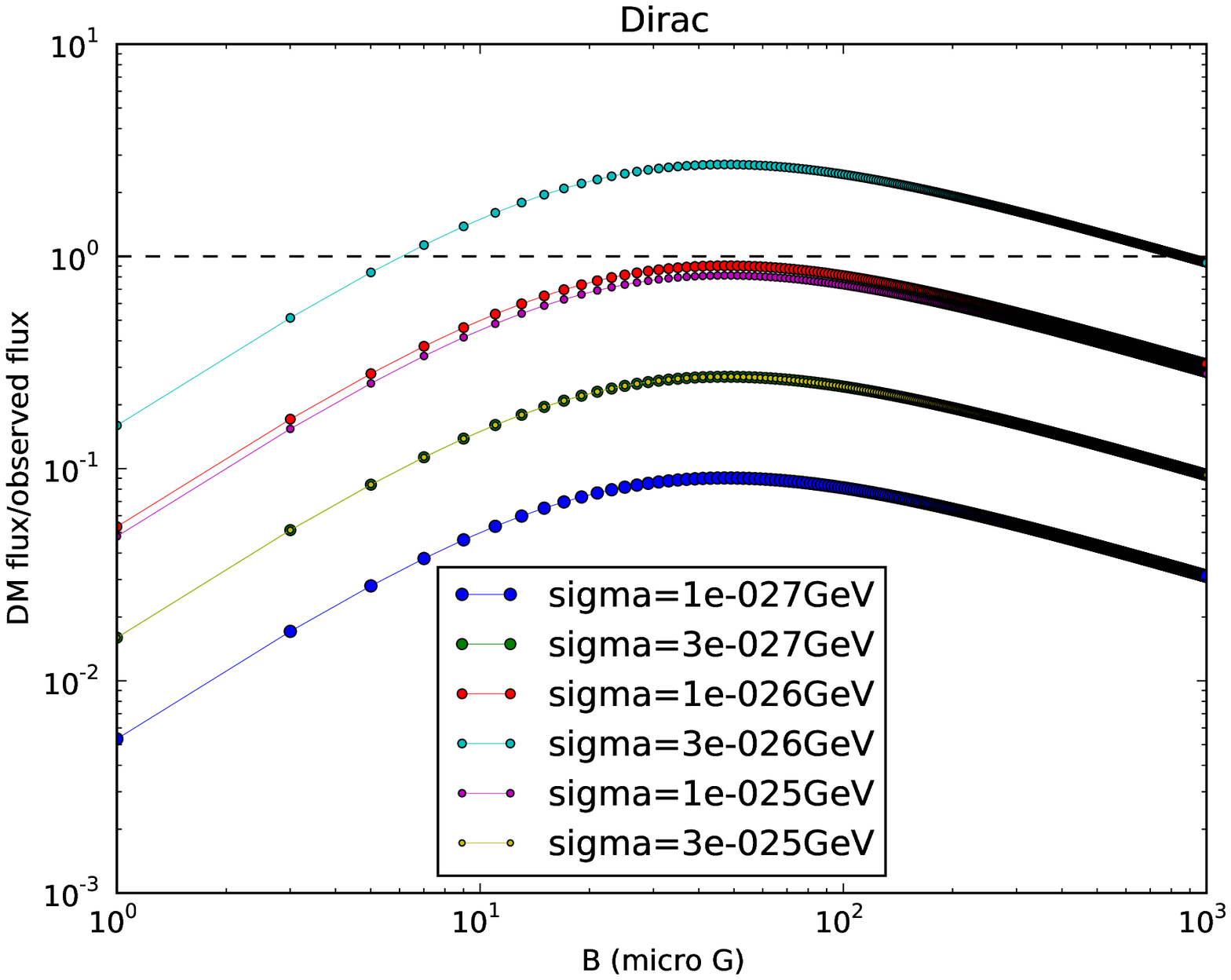}
	\caption{Ratio of the radio emission at 330 Mhz expected from dark matter annihilation in the inner galaxy ($10^{-5} \, sr$) to the observed radio emission at the same frequency versus the magnetic field. Here we have fixed the dark matter mass to 8 GeV and varied the annihilation cross section. We consider both  Majorana (left panel) and Dirac (right panel) annihilating particles.  All dark matter models above the dashed line (corresponding to a ratio equal to unity) are excluded since they would lead to a radio emission at 330 MHz that is greater than what has been observed in the inner part of the galaxy. Models 
	for which this ratio is greater than 0.1 are not necessarily excluded but they would change our understanding of the origin of the radio emission in the inner part of the galaxy (and probably the inferred measure of the magnetic field).  Models with ratio below 0.1 are likely to be viable although more careful studies are required. Here, we assume a NFW profile ($\gamma=1$). The curves for $\sigma v=3 \ 10^{-27}$ and $\sigma v=3 \ 10^{-25} \ \rm{cm^3/s}$ are identical because of the rescaling factor of the dark matter halo density that is required when $\sigma v=3 \ 10^{-25} \ \rm{cm^3/s}$.}
	\label{fig:sigma_ann}
\end{figure}

To illustrate this point, we display in Fig.~\ref{fig:sigma_ann}, the value of the S-wave cross section that is excluded by  radio observations of the inner part of the Milky Way for both Majorana and Dirac particles. When this cross section exceeds $3. \ 10^{-26} \ \rm{cm^3/s}$, we rescale the dark matter energy density in the halo by a factor $\xi = \frac{\sigma v}{3. \ 10^{-26} \ \rm{cm^3/s}}$. When the cross section is smaller than $3. \ 10^{-26} \ \rm{cm^3/s}$, we assume that the relic density is achieved either through another channel or another mechanism.

Therefore, unless the P-wave cross section is large enough in the early Universe or another mechanism/channel (that does not produce too many electrons) is proposed to explain the observed dark matter abundance at decoupling,  one can use synchrotron radiation prediction in the very inner part of the galactic centre to constrain the dark matter mass as a function of interactions. In other words, if dark matter is light, radio emission can be used as a guideline to build a viable dark matter model.

\begin{figure}
	\centering	\includegraphics[width=8cm]{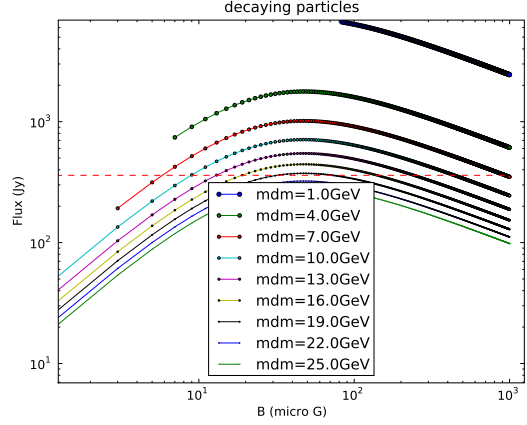}
	\caption{Flux expected in the Milky Way at 330 MHz for decaying dark matter candidates with masses equal to (from top to bottom) 1, 4, 7, 10, 13, 16, 19, 22, 25 GeV and a solid angle  of $\sim 10^{-5}$ sr. In this figure, we have taken a decay rate $\Gamma = 3 \ 10^{-26} \ \rm{s^{-1}}$ and a NFW profile ($\gamma=1$).}
	\label{fig:epsnud}
\end{figure}

Our results for the case of decaying dark matter are summarized in Fig.\ref{fig:epsnud}. Again, if the magnetic field is in between [20,100] $\mu$G, in the inner part of the galaxy, 
decaying dark matter candidates with a mass ranging from 1 to 25 GeV and a decay rate of $3 \ 10^{-26} \rm{s^{-1}}$ are expected to be ruled out. However, if the magnetic field is as large as 1 mG, particles with masses up to 7 GeV are excluded. Of course, one can rescale these results with the decay rate. It is interesting to note that the limits in the decaying case are quite similar to that in the annihilating scenario. This arises because the factor $r_s/d$ is only a few units in the galaxy unlike in clusters of galaxies, the resolution is good enough to be sensitive to the shape of the profile and the injection energy of the electrons is half that of the annihilating case.

In  the next section, we extend our calculations by adding the spatial diffusion of the electrons. Once again, we will focus on the very inner part of the galactic centre. 

\section{Synchrotron emission from dark matter annihilation or decay in presence of electron spatial diffusion \label{withdiffusion}}

In what follows, we adopt the semi-analytical approach detailed in Ref.~\cite{Delahaye:2007fr}. The function ${\cal{F}}_n$ that we computed analytically in the previous section is now replaced by a function $\tilde{I}$, referred to as the halo function. The latter includes the modelisation of the propagation of the electrons and positrons (it involves the characteristics of the diffusive halo and the diffusion coefficient). It also takes into account the energy losses. 

The larger the height of the diffusive halo, the higher electrons/positrons propagate leaving some evidences of emission at high latitude. Conversely if the height is too small it is difficult to distinguish the signal from dark matter to that of the background. In the following, we use a conservative estimate (referred to as the MED set of parameter, see Ref.~\cite{Delahaye:2007fr}), corresponding to a slab of $L=4$ kpc and a diffusive coefficient that has a spectral index of $\delta=0.7$ and a normalization of $K_0= 0.0112$ $\rm{kpc}^2/\rm{Myr}$. A larger slab would make the signal brighter at larger longitude and latitude.

Our main (simplistic) assumption is that the magnetic field is constant within $10^{-5}$ sr. In principle one should properly model its dependence  upon the distance but this is beyond the scope of this paper. Also we neglect the convection and reacceleration mechanisms while these may be important. At last, we assume an averaged electron number density to compute the Bremsstrahlung losses.

The surface brightness at a given location reads:

$$I_{\nu}(l,b) = < n(E)>_{(l,b)} \times \frac{E}{\nu} \times P(E) $$
where 
$$ < n(E)>_{(l,b)}  =  \frac{N_e}{{\eta_n} \ b(E)}   \ \int d s(l,b) \int d^3x \ Q_n \ \left(\frac{\rho(x)}{m_{dm}}\right)^n \times G(\odot,  \epsilon \leftarrow x,E_{inj}),$$ is the number of electrons of energy $E$, integrated over the line of sight, at a specific location (l,b). 

The term  $G(\odot,\epsilon \leftarrow x,E_{inj})$ represents the Green function that encodes the propagation of the electrons (spatial diffusion and losses) from their place of ``birth'' to a position $x$ and from an injection energy $E_{inj}=n \times m_{dm}/2$ down to an energy $E=\epsilon$.

\begin{figure}[h]
	\centering	\includegraphics[width=15cm]{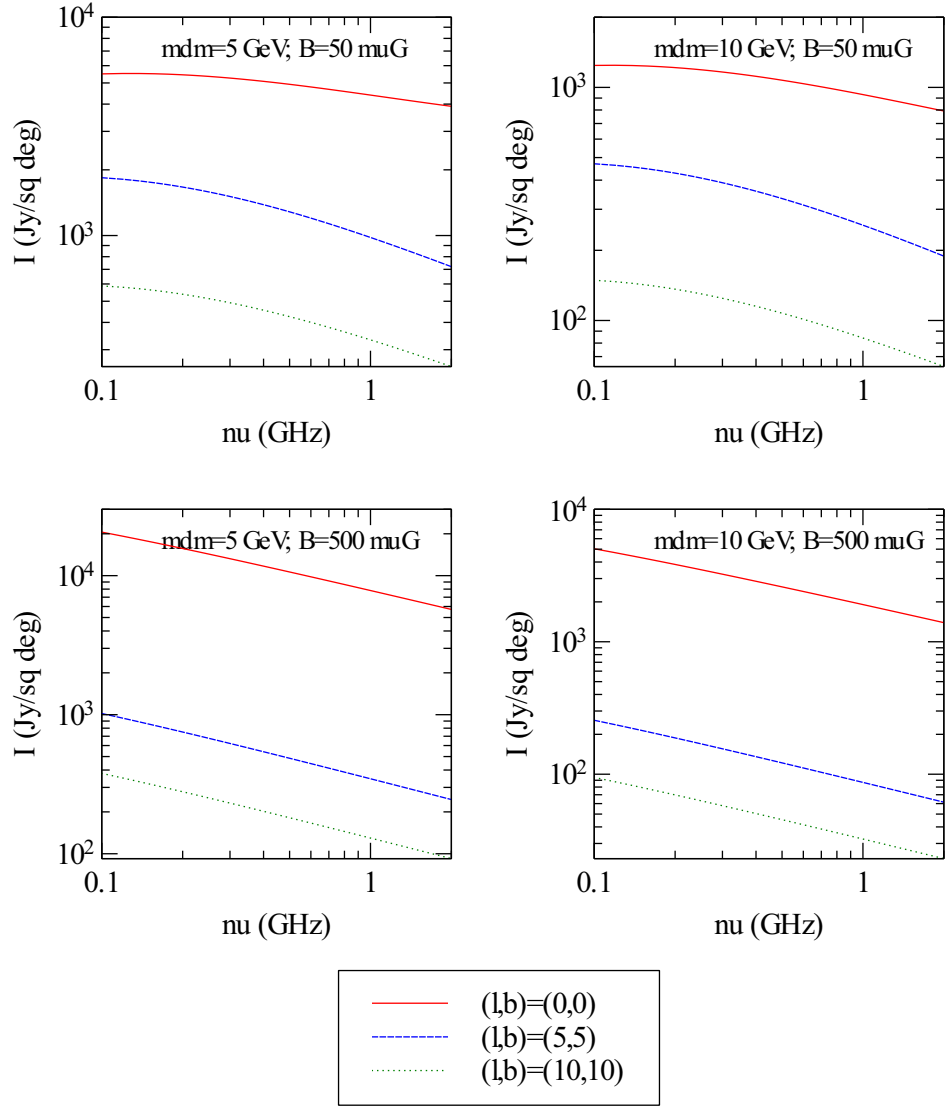}
	\caption{Surface brightness associated with annihilating self-conjugate dark matter candidates. We consider two masses in particular, namely $m_{dm}=5,10$ GeV, and three sets of longitude and latitude. For each dark matter mass, we investigate the effect of the magnetic field by taking $B=50 \mu$G and $500 \mu$ G. }
		\label{fig:ann_diffu}
\end{figure}

In Fig.\ref{fig:ann_diffu}, we show the expected surface brightness for self-conjugate dark matter candidates with a mass of 5 and 10 GeV respectively, using the MED set of parameters. We consider two values of the magnetic field, namely: $B=$ 50 and 500 $\mu$G. 

These figures show that when the magnetic field is not too strong, the electrons can diffuse out from the Galactic Centre. Thus $S_{\nu}$ at $l,b >0$ is not too suppressed at small frequencies with respect to the surface brightness in the centre. On the other hand, when the magnetic field is very strong, the electrons are confined within the Galactic Centre. Hence, the synchrotron emission is less significant outside the Galactic Centre than at $(l,b)=(0,0)$ and is stronger at low frequency than at high frequency (since the electrons can then lose their energy quickly where injection occurs).

\begin{figure}[h]
	\centering	\includegraphics[width=15cm]{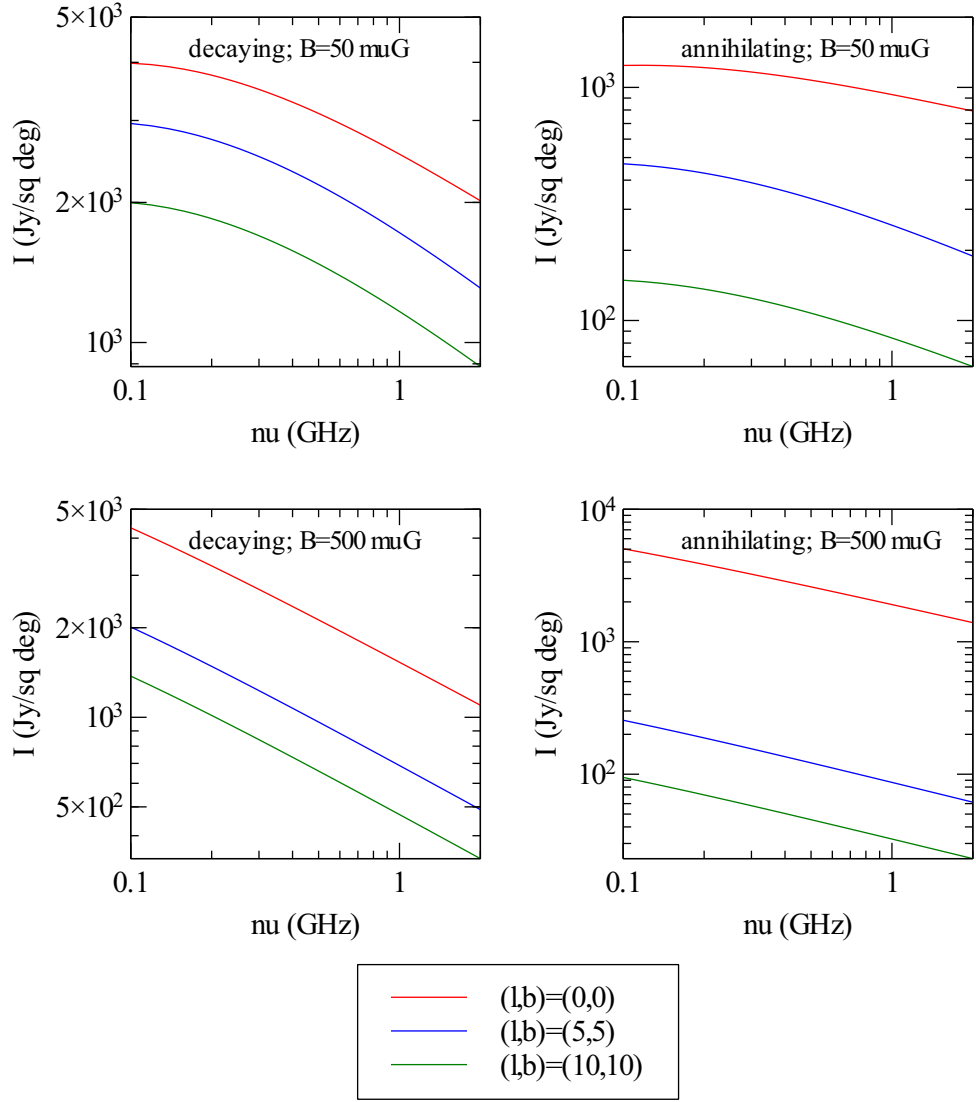}
	\caption{Surface brightness associated with both (self-conjugate) annihilating and decaying dark matter particles ($m_{dm}=10$ GeV) for three sets of longitude and latitude and $B=50 \mu$G and $500 \mu$ G. This figure shows that, for decaying particles, an increase in the magnetic field translate into a decrease in the surface bightness at high frequency. While, for annihilating dark matter, an increase of the magnetic field translates into a larger flux in the galactic centre at almost all frequencies because the electrons remain confined in the galactic centre (as this can be seen from the surface bightness at high latitude and longitude). The difference of behaviour between decaying and annihilating dark matter can be explained by the difference of injection energy of the electrons produced by the dark matter. In the decaying case, the injection energy is half that of the annihilating case and the bremsstrahlung losses are more important.}
	\label{fig:invert}
\end{figure}

In Tab.\ref{table}, we show the different values of the surface brightness for decaying and (self-conjugate) annihilating particles in the very inner part of the galactic centre.

\begin{table}[h]
\begin{center}
\begin{tabular}{|c|c|ccc|ccc|}
\hline
 &  &$S_{\nu, n=1}$ &$\times 10^{3} Jy/(deg)^2$   & &$S_{\nu, n=2}$ &$\times 10^3 Jy/(deg)^2$  &\\
 B $(\mu G)$ &$m_{dm} (\rm{GeV})$ &0.33 GHz &0.8 GHz &1.2GHz &0.33GHz &0.8 GHz &1.2 GHz\\
\hline
\hline
 50 &5 &7.26 &5.86 &5.23    &5.21 &4.57 &4.26\\
  &10  &3.43 &2.72 &2.39    &1.15 &0.978 &0.89 \\
  &15  &2.23 &1.76 &1.54      &0.48 &0.41 &0.37\\
  &20  &1.65 &1.30 &1.14    &0.26 &0.2 &0.199\\
\hline
 500  &5  &5.19 &3.43 &2.83     &12.7 &8.6 &7.2  \\
  &10     &2.57 &1.70 &1.40     &3.1 &2.1 &1.76 \\
  &15     &1.71 &1.13 &0.93     &1.36 &0.926 &0.772\\
  &20     &1.28 &0.84 &0.69     &0.763 &0.517 &0.431 \\
	\hline
\end{tabular}
\end{center}
\caption{Surface brightness expected for decaying and annihilating self-conjugate  dark matter particles in the galactic centre $(l,b)=(0,0)$ for two different values of the magnetic field ($B=50,500 \ \mu$G) and three frequencies. Here we take into account the propagation of the electrons. For non self-conjugate dark matter particles, one should divide these values by a factor two. }
\label{table}
\end{table}

We can now compare our results with observations. At 330 MHz, the surface brightness in Ref.\cite{LaRosa:2005ai} is smaller than $\sim$ 3-4 $10^3$ Jy per square degree. When $B=50 \mu$G,  we can therefore exclude annihilating candidates with $m_{dm} < 5$ GeV. These particles would indeed overproduce the synchrotron emission at this frequency in the inner galaxy. Self-conjugate particles up to 10 GeV seem also unlikely since they would also contribute significantly (about a third) to the synchrotron emission in the galactic centre. When $B=500 \mu$G, we can exclude annihilating particles up to $\sim$ 10GeV whether dark matter is self-conjugate or not. Self-conjugate particles up to 15 GeV should also be constrained by these observations. For decaying dark matter, the constraints are stronger when $B=50 \, \mu$G than when $B=500 \, \mu$ G. However, we can still exclude particles up to 10 GeV in both cases.

Note that we use a value of the magnetic field that is stronger than the one that is advocated in Ref.\cite{LaRosa:2005ai}. However, even though the surface brightness may be a bit smaller in the case of a weaker magnetic field, the dark matter candidate would still constitute a very large fraction of the synchrotron emission in the galactic centre.  Besides the emission at large latitude will be more significant,especially at small frequencies.

The behaviour of the surface brightness with the magnetic field is, in fact, in agreement the trend seen in Figs.\ref{fig:epsnua} and \ref{fig:epsnud} although by, adding spatial propagation, we have weakened our conclusions. For annihilating dark matter, an increase in the magnetic field actually increases the surface brightness in the galactic centre. The reason is that the electrons are injected at high energy and the synchrotron losses are comparable to the bremsstrahlung losses in the case of a large magnetic field. Thus, the electrons remain confined in the Galactic Centre and lose their energy without propagating very far. This does not happen in the decaying case because the electrons are injected at a smaller energy, where the losses are less important.

By focusing on the inner region $(l,b)=(0,0)$ where high magnetic field values are preferred, we found that decaying or (self and and non self-conjugate) annihilating dark matter particles with a mass up to 10 GeV are likely to be excluded if  $500 \ \mu \ \rm{G} > B > 50 \ \mu$ G.
This means that both the annihilation and decay rate must be smaller than the canonical value that we have taken here to avoid that light particles constitute a radio source in the Galactic Centre that is even more powerful than Sgr A. Yet, dark matter could still contribute to the radio emission in the galactic centre. This additional radio component will not exhibit any  variability and it may  be possible to trace the presence of such dark matter particles by using a multi-messengers analysis \cite{Colafrancesco:2005ji}.

Our results are somewhat more powerful than 
those in Ref.\cite{Borriello:2008gy} because we focused (as in \cite{bens}) on the very inner part of the galactic centre. Our findings seem nevertheless in agreement with the results displayed in Ref.\cite{Crocker:2010gy} for neutralino pair annihilations into electrons-positrons ($\chi \chi \rightarrow e^+ e^-$) when $m_{\chi} =$ 10 GeV.

\begin{figure}
	\centering	\includegraphics[width=16cm]{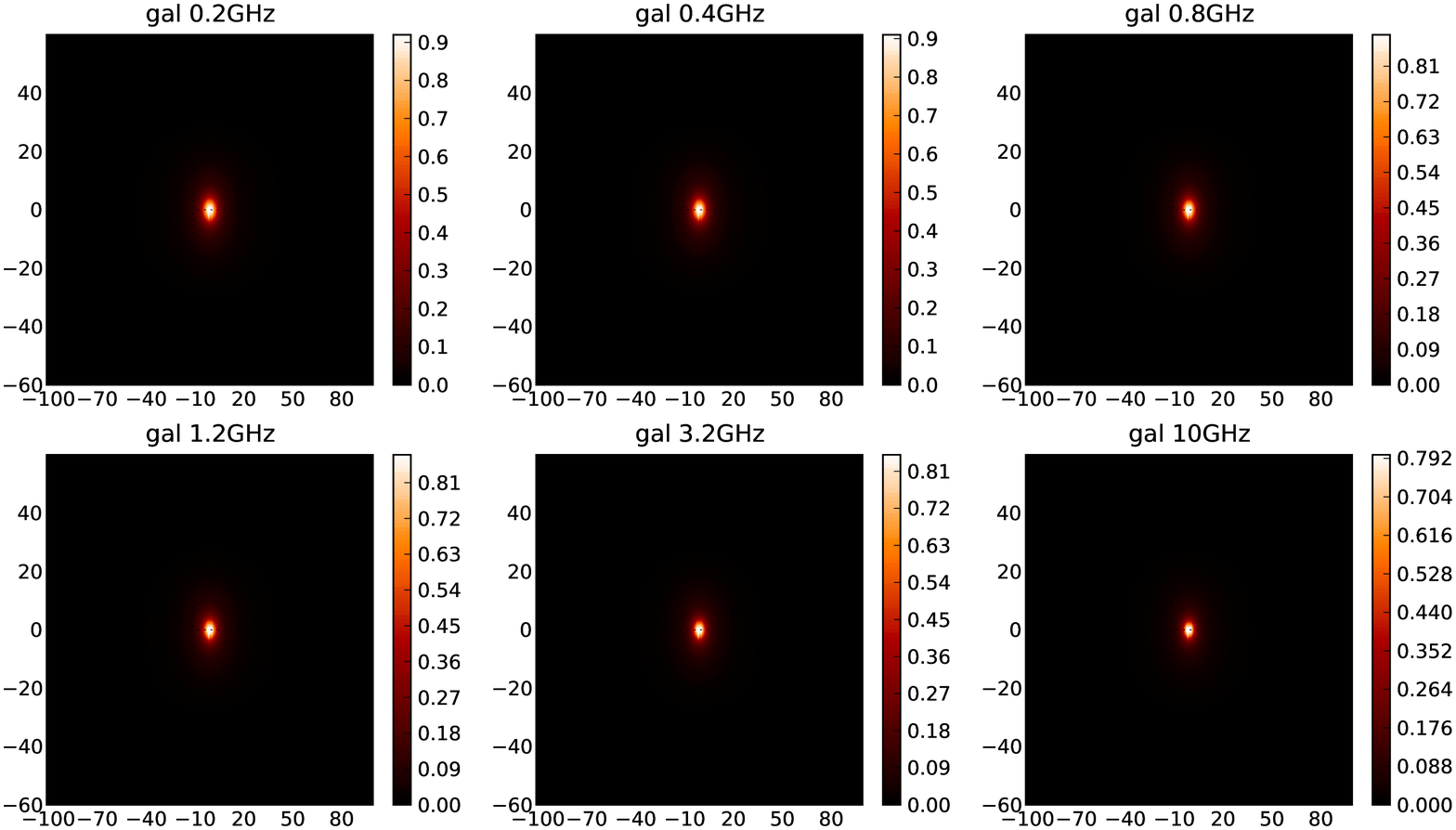}
	\caption{Map of the morphology of the synchrotron emission in the case of annihilating dark matter particles with $m_{dm}=10$ GeV. Here we consider a constant magnetic field of 100 $\mu$G. This is a too large value outside the galactic centre but this demonstates the effect of a strong magnetic field on the synchrotron emission in the galactic centre.}
	\label{fig:morpho_10gev_a}

\includegraphics[width=16cm]{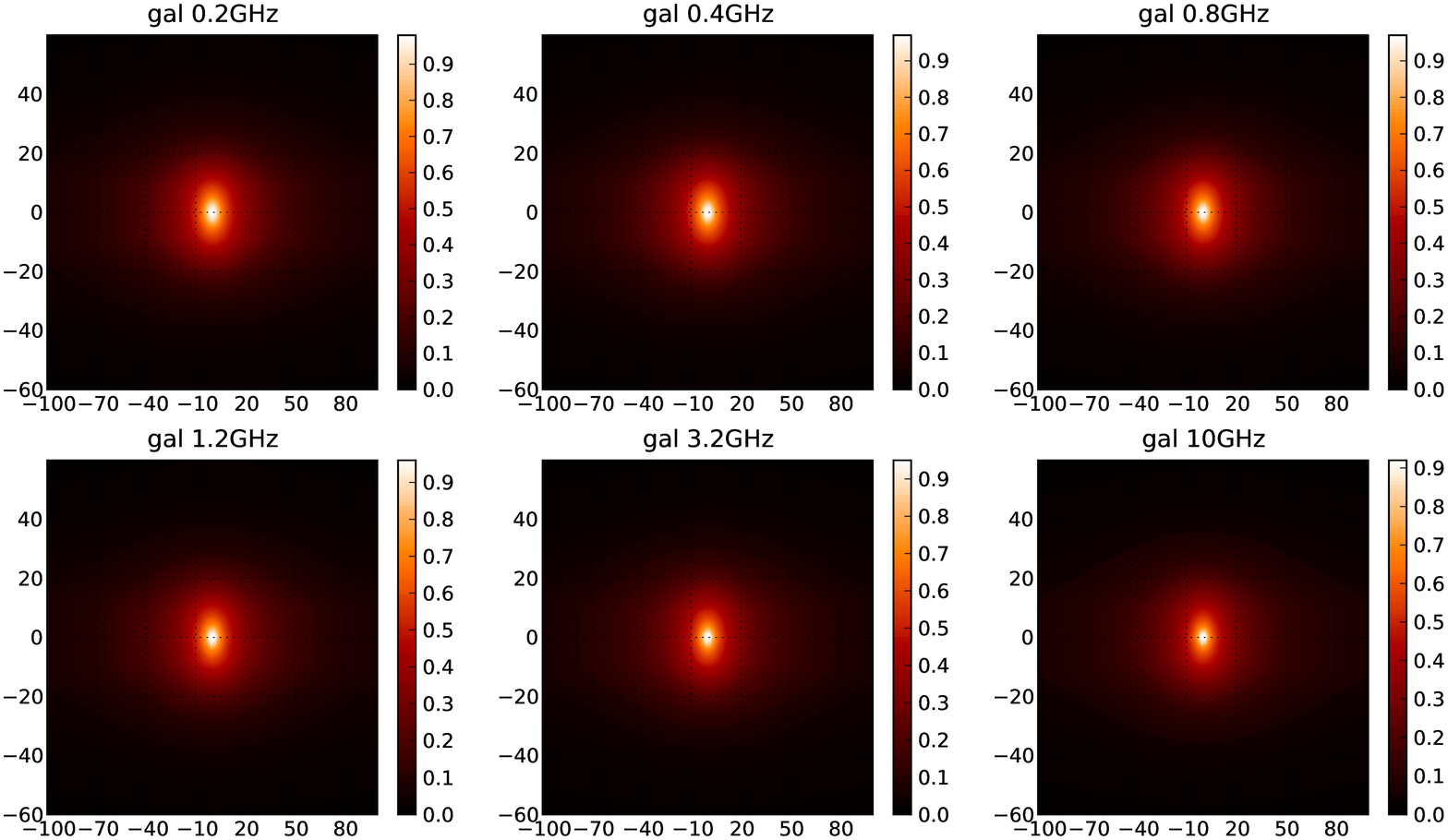}
	\caption{Map of the morphology of the synchrotron emission in the case of decaying DM with $m_{dm}=10$ GeV with a constant magnetic field of 100 $\mu$G.}
	\label{fig:morpho_10gev_d}
\end{figure}

In Figs.~\ref{fig:morpho_10gev_a} and \ref{fig:morpho_10gev_d}, we give the morphology of the radio emission that is expected for a $10$ GeV decaying and annihilating dark matter candidate and a magnetic field $B$ of  100 $\mu G$ for the following frequencies 0.2,0.4,0.8,1.2,3.2,10 GHZ. 
The difference between these two scenarios is striking.
As one can see, the signal is only bright at large latitude and longitude when the dark matter is decaying. If dark matter is annihilating, such a value of the magnetic field confines the dark matter in the centre and the emission is only bright in the very centre. Hence, this demonstrates that one can put severe constraints on \textit{annihilating} particles by using radio emission, but one should only focus on the inner galactic centre since the synchrotron emission  is likely to be too faint to be distinguished from the background and foreground sources far away from the galactic centre. Of course, taking a magnetic field as large as 100 $\mu$G in the whole galaxy is not realistic. However, even in presence of a smaller magnetic field, one cannot expect a bright emission at large latitude and longitude if dark matter since the power radiated will be anyway smaller. Hence the best place to constrain this scenario is actually the galactic centre. 

The chances of discovering or constraining \textit{decaying} dark matter using radio emisison are much greater. As one can see from Fig.\ref{fig:morpho_10gev_d}, the observation of a diffuse radio emission at these very low frequencies, with an elliptical shape, would indicate the presence of decaying dark matter candidates within the Milky Way dark halo. Although this map was obtained with a very large magnetic field, the morphology of the emission does not change much if we consider a smaller magnetic field because the Bremsstrahlung losses were dominant at these injection energies. Of course, this analysis remain quite simplistic and proper dedicated studies are required to make accurate predictions.

\section{Conclusion}

Recent claims from direct detection experiments have encouraged theoreticians to propose dark matter candidates in the 1-10 GeV range. Although the models which have been recently proposed are compatible with recent particle physics constraints, these studies do not take into account the astrophysical constraints that were first derived in Ref.~\cite{bens} on light (1-10 GeV) dark matter candidates. Yet, as we pointed out in 2002, dark matter particles in this mass range are likely to overproduce the radio emission in the galactic centre and yield a visible signature in Coma. Therefore these constraints should be carefully addressed.

In this paper, we  update and extend our previous calculations. We derive the synchrotron spectrum that is to be expected in the galactic centre (with and without spatial propagation) and in Coma. We also give the morphology of the emission for 5,10 GeV. We show that the flux associated with annihilating and decaying particles (with a mass in the 1-10 GeV range) would constitute a large fraction (if not all) the synchrotron emission in the galactic centre if their annihilation rate into electrons and positrons is about $3 \ 10^{-26} \rm{cm^3 s^{-1}}$ or their two-body decay rate into $e^+ e^-$ is about $\Gamma=3 \ 10^{-26} \rm{s^{-1}}$. 

Therefore, not only have these constraints  the ability to exclude many models which have been proposed to explain recent signals in dark matter direct detection experiments but they also suggest that if dark matter is annihilating and has a mass in the $[1-10]$ GeV range, it must have a S-wave suppressed pair annihilation cross section into electrons with respect to the canonical value. This is particularly true if the dark matter consists of self-conjugate annihilating particles. Also all channels producing electrons must be suppressed. Hence, if  dark matter is really made of thermal annihilating particles in this mass range, the first question that one will have to answer in order to build a model is: what is the mechanism which provides the dark matter with an acceptable relic density?

\section{Acknowledgment}
We are grateful to D. Albornoz-V\'asquez, G. B\'elanger and T. Delahaye for valuable comments. The collaboration between C.B. and J. S. is supported by a CNRS PICS.
\bibliography{10GeV}

\end{document}